\documentclass[]{interact}

\usepackage{epstopdf}
\usepackage[caption=false]{subfig}

\usepackage{caption}
\usepackage[dvipsnames]{xcolor}

\usepackage[numbers,sort&compress]{natbib}
\bibpunct[, ]{[}{]}{,}{n}{,}{,}
\renewcommand\bibfont{\fontsize{10}{12}\selectfont}

\theoremstyle{plain}

\theoremstyle{definition}

\theoremstyle{remark}

\usepackage{ wasysym }
\newcommand{\dd}{\mathrm{d}}

\begin{document}


\title{Effect of hydrogen addition on the consumption speed of lean premixed laminar methane flames exposed to combined strain and heat loss}

\author{
\name{Alex M. Garcia \textsuperscript{a,b}\thanks{Email: alex.garcia@tum.de}, Sophie Le Bras \textsuperscript{a}, and Wolfgang Polifke\textsuperscript{b}}
\affil{\textsuperscript{a}Siemens Industry Software N.V, 3001, Leuven, Belgium;\\ \textsuperscript{b}Technical University of Munich, School of Engineering \& Design, D-85747 Garching, Germany}
}

\maketitle

\begin{abstract}

This study presents a numerical analysis of the impact of hydrogen addition on the consumption speed of premixed lean methane-air laminar flames exposed to combined strain and heat loss. Equivalence ratios of 0.9, 0.7, and 0.5 with fuel mixture composition ranging from pure methane to pure hydrogen are considered to cover a wide range of conditions in the lean region. The 1-D asymmetric counter-flow premixed laminar flame (aCFPF) with heat loss on the product side is considered as a flamelet configuration that represents an elementary unit of a turbulent flame and the consumption speed is used to characterize the effect of strain and heat loss. Due to the ambiguity in the definition of the consumption speed of multi-component mixtures, two definitions are compared. The first definition is based on a weighted combination of the consumption rate of the fuel species and the second one is based in the global heat release rate. The definition of the consumption speed based on the heat release results in lower values of the stretched flame speed and even an opposite response to strain rate for some methane-hydrogen-air mixtures compared to the definition based on the fuel consumption. Strain rate leads to an increase of the flame speed for the lean methane-hydrogen mixtures, reaching a maximum value after which the flame speed decreases with strain rate. Heat loss decreases the stretched flame speed and leads to a sooner extinction of the flamelet due to combined strain and heat loss. Hydrogen addition and equivalence ratio significantly impact the maximum consumption speed and the flame response to combined strain rate and heat loss. The effect of hydrogen on the thermo-diffusive properties of the mixture, characterized by the Zel’dovich number and the effective Lewis number, are also analyzed and related to the effect on the consumption speed. Two definitions of the Lewis number of the multi-component fuel mixture are evaluated against the results from the aCFPF.

\end{abstract}


\section{Introduction}

Hydrogen addition to hydrocarbon-air flames enhances the chemical reactivity of the combustion process leading to higher flames speeds and wider flammability limits \cite{YuLaw86}. The blending of hydrogen with conventional hydrocarbon gas fuels has received increased attention in order to reduce CO$_2$ emissions, especially in the sector of large-scale power generation, where the higher reactivity of the hydrogen blends makes possible ultra-lean premixed combustion systems without a negative impact on combustion efficiency and emissions of carbon monoxide and unburned hydrocarbon, even though it may be more susceptible to thermo-acoustic instabilities \cite{BeitaTalib21}. 

For situations in which the laminar flame thickness is smaller than the turbulent length scales, a premixed turbulent flame can be considered as a statistical ensemble of premixed laminar flames known as "flamelets", which are corrugated and wrinkled by the turbulent flow field \cite{PoinsVeyna12a}. Thus, the propagation speed of the turbulent flame is a function of the wrinkled flame surface area and the local flame speed of the flamelets. In many applications, such as gas turbines, the flamelets are exposed to high levels of flame surface deformation or stretch, and the turbulent flame speed may vary because the local laminar flame speed varies with stretch. Moreover, under high enough stretch flamelets may be extinguished \cite{KarloDenni53, Rogg88}, which has a big effect on the global propagation speed of the turbulent flame. 

The effect of strain rate is highly related to the thermo-diffusive  properties of the reactant mixture. If the thermal diffusion is lower than the mass diffusion (Lewis number Le $<$ 1), the flame speed increases with strain rate. For pure hydrogen, thermo-diffusive effects result in large changes in the local flame speed, that can be more than twice larger than the unstretched laminar flame for positive local stretch conditions along a turbulent flame front \cite{ChenIm00}. Therefore, the addition of hydrogen to a methane-air flame reduces its thermo-diffusive stability causing the flame speed to increase with positive stretch \cite{Chen09a,OkafoHayak14}. 

Considering the increase of flame speed with positive stretch is crucial for the understanding and modeling of methane-hydrogen premixed turbulent flames. Various fuel mixtures can have substantially different turbulent flame speeds, despite having the same unstretched laminar flame speed and turbulence intensity \cite{KarpoSokol61, KidoNakah02}, due to their different response to stretch. Experiments with methane-hydrogen mixtures have shown that hydrogen addition triggers the transition from V to M-shape flame in non-adiabatic turbulent flames, for mixtures with the same unstretched laminar flame speed \cite{GuibeDurox15a, TaamaShanb16}. 

Leading point theories suggest that the premixed turbulent flame speed is controlled by the characteristics of the flamelets ahead of the flame brush that advance farthest into the unburned mixture, called leading points or leading edge \cite{KuzneSabel90,LipatChomi05}. The flamelets at the leading edge are positive stretched (strain and curvature); therefore, the local flame speed will increase for mixtures with Le $<$ 1. The increase in the flame speed will drive the leading edge to propagate further into the unburned reactants, further increasing the stretch experienced by these flamelets and, therefore, the local flame speed \cite{VenkaMarsh11}. The process of increasing flame speed could continue until the leading point reaches the maximum stretched flame speed of the mixture. This makes the maximum stretched flame speed an important parameter in turbulent flame combustion, where the flame response to stretch governs the behavior of the reactive mixture \cite{VenkaMarsh11,SalusBergt15}. Accordingly, the stretched flame speed and the extinction strain rate can be used to characterise the flame shape, flow structure and thermo-acoustics stability of different fuel mixtures in various combustion systems \cite{HongShanb15, MichaShanb17, AltaySpeth09, SpethGhoni09, ShanbSanus16}.

Three effects can induce stretch in a flame: the curvature of the flame surface, the movement of a curved part of the flame surface, and the aerodynamic strain produced by flow non-uniformity along the flame surface \cite{Law89}. These stretch effects could be either positive or negative and they have different effects on the flame chemistry \cite{ChenBottl21}. In a turbulent flame, the positive and negative curvature tend to balance each other when they are averaged along the flame front (mean value of curvature near zero), while the strain rate tends to be overall positive \cite{HaworPoins92,EchekChen96}. Therefore, the present work focuses on the stretch induced by positive strain rate. However, the effect of hydrogen addition on the behavior of the stretch induced by curvature may differ from the one induced by strain rate \cite{BottlSchol21} and should be also studied. \color{black}

Under non-adiabatic conditions, flamelets are more sensitive to being extinguished by stretch effects, and both the flame speed and the extinction strain rate decrease due to heat loss. Different approaches can be used to consider non-adiabatic conditions, such a reducing the energy at the boundary of an adiabatic flamelet, introducing heat loss as a sink term in the flamelet energy equation, or reducing the enthalpy of the reactant mixture. However, the asymmetric counterflow flame (fresh-to-burnt) with reduced enthalpy products is a more suitable representation of the stretched flamelets in a turbulent flame confined with non-adiabatic walls, where reduced temperature products flow back and interact with the flame. Therefore, this flamelet configuration is an excellent building block for turbulent combustion models that consider the combined effect of stretch by strain rate and heat loss \cite{Tay-WKomar09,NassiPampa21,TangRaman21,KutkaAmato22}. 

The effect of heat loss in the asymmetric counterflow premixed flame has been studied for hydrocarbon-air mixtures \cite{Dixon96,Dixon06,LibbyLinn83,LibbyWilli83, DarabCande88,Dixon91,Tay-WScarp17}. Most of these studies use asymptotic analysis, giving qualitative results of the effect of various parameters without specifying a particular reactant mixture. In the case of methane-hydrogen mixtures, the response to strain rate has been studied mainly on outwardly propagating spherical flames \cite{HuangZhang06,HuHuang09,Chen09a,OkafoHayak14, BouveHalte13,SankaIm06} without considering the effect of heat loss. Therefore, the current study presents a numerical analysis of the impact of hydrogen addition on the response of premixed lean methane-air laminar flames to positive strain rate and heat loss. Special attention is paid to the maximum stretched flame speed and the extinction by combined strain rate and heat loss effects. Moreover, two definitions of consumption speed valid for multi-component fuel mixtures are compared to observe their capability to describe flamelets with the thermo-diffusive properties of methane-hydrogen mixtures under the strain rate and heat loss conditions that may be experienced in a non-adiabatic turbulent flame.

\begin{figure}[h]
    \centering\includegraphics[trim=0 280 600 0, clip, width=0.55\linewidth]{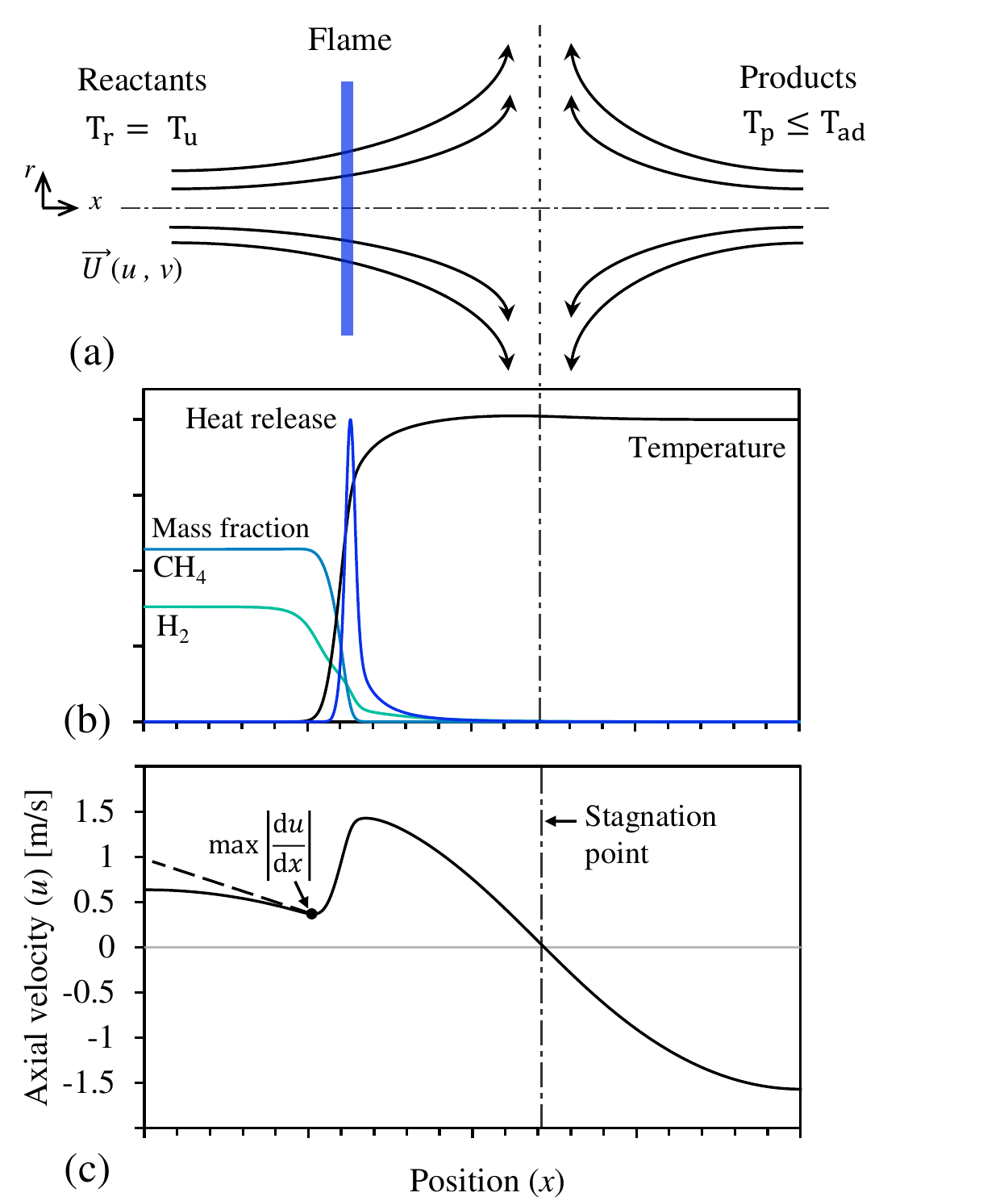}
    \caption{(a) Schematic of the counter flow premixed laminar flame, (b) characteristic profiles of temperature, heat release and fuel mass fraction, and (c) characteristic axial velocity profile.}
    \label{fig_CFPF_scheme} 
\end{figure}


\section{Methodology}
The asymmetric counterflow premixed flame (aCFPF) configuration is used to study the effect of positive strain and heat loss on the flame speed. A schematic of this configuration is shown in Fig.~\ref{fig_CFPF_scheme}(a). It consists in two opposing streams of reactants and products, which create a stagnation point flow with a planar flame located at the reactant side of the stagnation point. The subscripts $\cdot_r$ and $\cdot_p$ are used to denote reactant and product quantities, respectively. The typical evolution of the profiles of heat release, temperature, and fuel mass fraction are shown in Fig. 1(b).

\subsection{Strain rate}

The characteristic strain rate of the counterflow flame is defined as the maximum gradient of the axial velocity, $\kappa = \mathrm{max}(|du/dx|)$, in the hydrodynamic zone upstream of the flame position, as shown in Fig.~\ref{fig_CFPF_scheme}(c)). This definition prevents $\kappa$ from being affected by the flame response  \cite{Law89,Rogg88}. The mass flow rates of reactants and products are related by a condition of equal momentum, $\rho_r U_r^2 = \rho_p U_p^2$ \cite{Rogg88}, where $\rho$ and $U$ are the density and the  axial velocity magnitude, respectively. Thus, as the mass flow rate of the reactants increases, the flame is more strained by the whole velocity field. While $\kappa$ increases, the flame (reaction zone) moves toward a position closer to the stagnation point where the local flame speed balances the local axial velocity. 
The strain rate is expressed in dimensionless form using the Karlovitz number Ka, which is the ratio between the chemical time scale and the flow time scale. For laminar flames, Ka is given as:

\begin{equation}
\mathrm{Ka} = \frac{\delta_L^0}{S_L^0} \kappa,
\label{Eq_Ka} 
\end{equation}

\noindent where $\delta_L^0$ is the diffusive laminar flame thickness, defined as the ratio of the thermal diffusivity of the reactants and the unstretched laminar flame speed $S_L^0$.

\subsection{Heat loss}


Due to flow fluctuations in a premixed turbulent flame some products flow back and interact with the flame. Under non-adiabatic conditions, these products may lose heat by convection before reaching the flame again. In that sense, heat loss is imposed on the product side of the aCFPF, and a heat loss coefficient may be defined as \cite{Tay-WZellh15}:

\begin{equation}
\beta = \frac{T_p - T_r}{T_{ad} - T_r},
\label{Eq_Beta} 
\end{equation}

\noindent where $T_r$ is the temperature of the reactants, $T_p$ is the temperature of the product stream, and $T_{ad}$ is the temperature of the products under adiabatic conditions. In order to model non-adiabatic conditions, the value of $T_p$ is progressively decreased from $T_{ad}$, leading to $\beta<1$. Radiation heat loss is not included in this numerical study since its effects can be neglected for mixtures not close to their flammability limits and without CO$_2$ or H$_2$O dilution \cite{Dixon06,ChenGou10,Egolf94}.

\begin{figure}[h]
    \centering\includegraphics[trim=10 290 10 295, clip, width=1\linewidth]{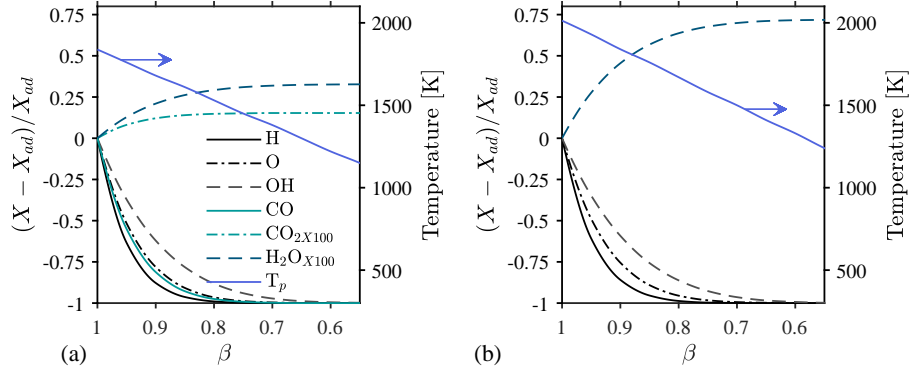}
    \caption{Example of the product stream composition variation with temperature, for (a) pure methane and (b) pure hydrogen, with $\phi=0.7$ and $T_r$=300 K. $X$: mole fraction at T$_p$, $X_{ad}$: mole fraction at T$_{ad}$.}
    \label{fig_xprod} 
\end{figure}

The composition of the products is also changed according to the heat loss condition. For the adiabatic case, the composition corresponds to the equilibrium composition at constant enthalpy and pressure of the reactant mixture (i.e. equilibrium at the adiabatic flame temperature). For $T_p<T_{ad}$, the composition changes to be in chemical equilibrium at the new temperature. As an example, Fig.~\ref{fig_xprod} shows the evolution of the molar concentration of some species in the products as a function of the coefficient $\beta$, for pure methane and pure hydrogen. While $T_p$ decreases from the adiabatic flame temperature, the equilibrium concentrations of the intermediate species and free radicals drop sharply with temperature. When this is not considered, an artificial diffusion of radicals from the products side of the stagnation point to the reaction zone may occur. 

\subsection{Consumption speed}

The consumption speed represents the rate at which the flame consumes the reactants and is a suitable measure of flame speed under a variety of stretched conditions \cite{PoinsVeyna12a,ImChen00}. The consumption speed, $S_{c_F}$, is obtained from the integration of the fuel mass conservation equation along the axis normal to the flame front, as \cite{PoinsVeyna12a}:

\begin{equation}
    S_{cF} \equiv -\frac{1}{\rho_u \displaystyle Y_{f,u}}\int_{-\infty}^{\infty} \dot{\omega}_{f}\dd x,
\label{Eq_Sc_TNC} 
\end{equation}

\noindent where $Y_f$ and $\dot{\omega}_f$ are the mass fraction and the volumetric mass production rate of the fuel, respectively. Equation~\eqref{Eq_Sc_TNC} leads to a non-unique definition for multi-component fuel mixtures. For an unstretched laminar flame (Ka = 0), the consumption speed corresponds to the the value of $S_L^0$, irrespective of the species chosen for its definition. However, the definition based on different species may lead to very different consumption speeds for a stretched flame if they have highly different diffusion properties. Therefore a suitable definition must be chosen for multi-component fuel mixtures such as CH$_4$-H$_2$. In the present work the consumption speed for the multi-component fuel is defined by a weighted combination of the $N_f$ species in the fuel mixture as:
\color{black}
\begin{equation}
    S_{cF} \equiv \frac{\displaystyle \sum_{k=1}^{N_f}{ \eta_k \int_{-\infty}^{\infty} \dot{\omega}_k}\dd x}{\rho_u \displaystyle \sum_{k=1}^{N_f}{\eta_k (Y_{k,b} - Y_{k,u})}},
\label{Eq_Sc_Fuel} 
\end{equation}

\noindent where the subscripts $\cdot_u$ and $\cdot_b$ denote the unburned and burned sides of the flame. Different values of the weighting coefficient $\eta_k$ have been used for some CH$_4$-H$_2$ mixtures \cite{SankaIm06,VanceShosh20}. In the present work, the species mass fraction in the fuel mixture is used as a weighting factor, $\eta_k=Y_{k,u}/Y_{f,u}$. This is done in order to match the one-component fuel definition (Eq.~\eqref{Eq_Sc_TNC}), mainly in the case of pure CH$_4$, where H$_2$ is still present in the reaction as an intermediate species.

An alternative way to define the consumption speed is based on the heat release rate by integrate the energy conservation equation, yielding:

\begin{equation}
    S_{c \dot{Q}} \equiv \frac{\displaystyle \int_{-\infty}^{\infty} \dot{Q}  \, \dd x}{\rho_u c_p (T_b - T_u)},
    \label{Eq_Sc_HR1} 
\end{equation}

\noindent where $\dot{Q}$ is the total volumetric heat release and $c_p$ is the mass-specific heat. The definition based on the heat release eliminates the need to choose a species or weighting factor to define the flame speed. Thus, it can be used for arbitrary fuel mixture without any change. However, for the present analysis, where heat loss at the product side is considered, the consumption speed is not well-defined by Eq.~\eqref{Eq_Sc_HR1}. This is due to the fact that the flame temperature $T_b$ changes as a result of heat loss, which is not considered when the equation of energy conservation is integrated. Equation~\eqref{Eq_Sc_HR1} could be modified based on the idea that all the sensible enthalpy (denominator in Eq.~\eqref{Eq_Sc_HR1}) comes from the enthalpy of formation (chemical enthalpy) of all the species involved in the oxidation reaction, leading to:  

\begin{equation}
    S_{c { \dot{Q}}} \equiv \frac{\displaystyle \int_{-\infty}^{\infty} \dot{Q}  \, \dd x}{\rho_u \displaystyle \sum_{k=1}^{N}{\Delta h_{f, k}^0 (Y_{k,b} - Y_{k,u})}}.
    \label{Eq_Sc_HR2} 
\end{equation}

In the following, the definitions of the consumption speed given in Eqs.~\eqref{Eq_Sc_Fuel} and \eqref{Eq_Sc_HR2} are considered, and their respective influence on the response of the flames to strain and heat loss is discussed. 
\subsection{Effective Lewis number}

As indicated previously, the response of the laminar flame to strain highly depends on the thermo-diffusive properties of the mixture: thermal diffusivity and mass diffusivity of the various reactant species. The Lewis number Le relates the thermal and mass diffusion as follows:

\begin{equation}
    \mathrm{Le}_i \equiv \frac{D_{th}}{D_{i,N_2}},
    \label{Eq_Le_i} 
\end{equation}

\noindent where $D_{th}$ is the thermal diffusion of the fuel-air mixture, and $D_{i,N_2}$ is the bi-molecular diffusivity of the species $i^{th}$ and nitrogen N$_2$. $D_{i,N_2}$ is normally taken as the characteristic diffusivity of the $i^{th}$-species because N$_2$ is the most abundant species in the fuel-air mixture.

An effective Lewis number for the reactant mixture is necessary to analyze the mixture as a whole. However, there is no general agreement on the most appropriate definition of this quantity. Different definitions have been proposed in the literature to calculate the Lewis number of various fuel mixtures \cite{LawJomaa05, MuppaNakah09, DinkeManic11, BouveHalte13}. The two most commonly used definitions of the Lewis number for the fuel mixture CH$_4$-H$_2$ are compared in the present work. The definition based on the volume fraction expresses the fuel Lewis number as \cite{MuppaNakah09}:

\begin{equation}
\mathrm{Le}_{fuel~V} = X_{CH_4}\mathrm{Le}_{CH_4} + X_{H_2}\mathrm{Le}_{H_2},
\label{Eq_LeV_fuel} 
\end{equation}

\noindent while the definition based on the species mass diffusion expresses the fuel Lewis number as \cite{DinkeManic11}:

\begin{equation}
\mathrm{Le}_{fuel~D} = \frac{D_{th}}{X_{CH_4}D_{CH_4,N_2} + X_{H_2}D_{H_2,N_2} } =  \frac{1}{X_{CH_4}/\mathrm{Le}_{CH_4} + X_{H_2}/\mathrm{Le}_{H_2}},
\label{Eq_LeD_fuel} 
\end{equation}

\noindent where $X_i$ is the volume fraction of the species $i^{th}$ in the fuel mixture, and Le$_i$ is the Lewis number of the fuel species $i^{th}$ in the fuel-air mixture (CH$_4$-H$_2$-air), using Eq.~\eqref{Eq_Le_i}. \color{black}

In a lean mixture, the fuel is the reactant species in deficit and oxygen is the reactant species in excess. Therefore, the effective Lewis number of a lean fuel-air mixture is given by \cite{BechtMatal01}:

\begin{equation}
\mathrm{Le}_{\mathrm{eff}} = \frac{\mathrm{Le}_{O_2,stq}+A \mathrm{Le}_{fuel,stq}}{1 + A}, \quad \mathrm{with} \quad A= 1+\mathrm{Ze}(1/\phi -1)
\label{Eq_Le_eff} 
\end{equation}

\noindent where Le$_{O_2}$ is the Lewis number of the oxygen and $\phi$ is the equivalence ratio of the lean mixture. The subscript $\cdot_{stq}$ indicates that the Le numbers correspond to the stoichiometric fuel-air mixture; otherwise, the effect of the equivalence ratio described by Eq.~\eqref{Eq_Le_eff} will be wrong. The Zel'dovich number is given by Ze $= [E_a (T_{ad}-T_u)]/(R T_{ad}^2 )$, with $R$ and $E_a$ being the universal gas constant and the overall activation energy, respectively. The overall activation energy represents the sensitivity of the laminar flame speed to flame temperature variations \cite{SunSung99}.



\subsection{Numerical model}

The aCFPF is simulated using the software Cantera \cite{GoodwMoffa18}. A discretized version of the one-dimensional conservation equations that govern momentum, energy, and species mass transport are solved along the stagnation streamline (central axis) using the general formulation derived by Kee et al. \cite{KeeColtr17}. An adaptive gridding scheme is used and the burner separation distance is set to 300 times the laminar flame thickness to reduce its effects on the strain rate. The oxidation of the reacting mixture is described using the detailed chemical kinetic mechanism GRI 3.0 \cite{SmithGolde99}, and full multi-component mass diffusion and Soret effect are considered. The reference pressure for the analysis is 101.3 kPa and the temperature of the reactants is kept constant at $T_r$ = 300 K. The methane-hydrogen mixtures evaluated range from pure CH$_4$ to pure H$_2$ with increments of 20$\%$ H$_2$ by volume, and referred in terms of the hydrogen mole fraction in the fuel mixture $X_{H_2}$. The equivalence ratios covered are $\phi$ = 0.9, 0.7, and 0.5. Finally, the strain rate for each mixture varies from a value corresponding to $U_r \simeq S_L^0$ to a value for which $S_c$ is zero or changes only very slightly after a considerable increase in the strain rate.

The freely propagating laminar flame is simulated to compute the unstretched laminar flame speed $S_L^0$ and the overall activation energy for each CH$_4$-H$_2$-air mixture, the latter by varying the N$_2$ concentration in the oxidant at constant equivalence ratio \cite{SunSung99}.


\section{Results and discussion}


\subsection{Consumption speed for adiabatic strained laminar flame}

\begin{figure}[h]
    \minipage{0.5\textwidth}
      \includegraphics[trim=165 185 165 185, clip, width=\linewidth]{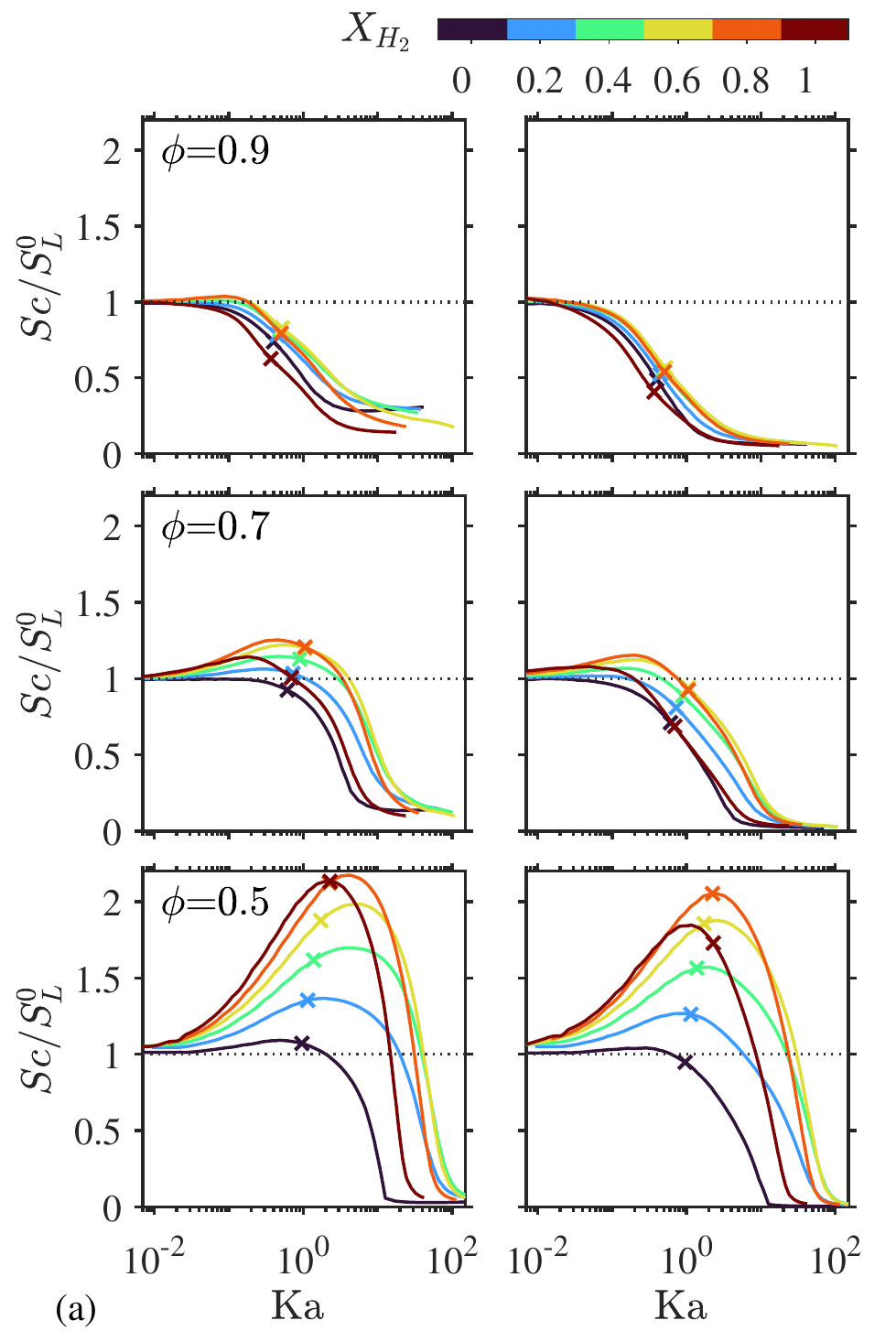}
    \endminipage
    \minipage{0.5\textwidth}
      \includegraphics[trim=165 185 165 185, clip, width=\linewidth]{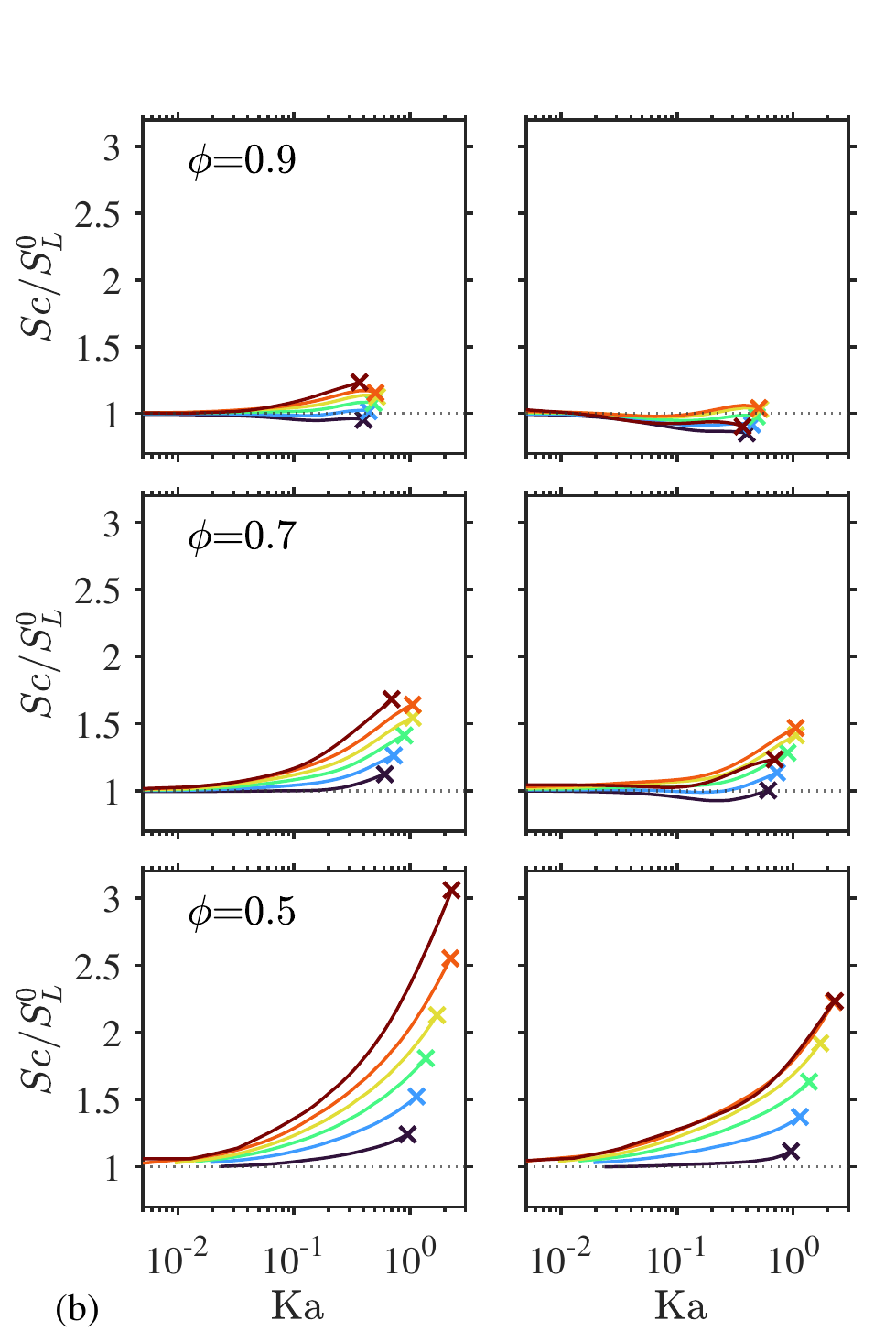}
    \endminipage
    \caption{Response of CH$_4$-H$_2$ laminar flame to strain (a) in the adiabatic aCFPF and (b) in the twinCFPF. Consumption speed based on fuel consumption S$_{cF}$ (left), and based on heat release rate S$_{c{\dot{Q}}}$ (right). The extinction strain rate from the twinCFPF is shown by the marker ($\times$) on top of the aCFPF results for comparison.}
    \label{plot_SRvSc_Ad} 
\end{figure}

In order to easily discern the effect of combined strain rate and heat loss on the laminar flame response, the impact of hydrogen on the consumption speed under adiabatic conditions is discussed first in this section. Figure~\ref{plot_SRvSc_Ad}(a) presents the evolution of the consumption speed in the aCFPF as a function of the strain rate when the temperature of the products is equal to $T_{ad}$ ($\beta$=1). The results from both definitions of consumption speed are presented side by side.

Overall, the results obtained for the heat-release-based consumption speed S$_{c{\dot{Q}}}$ show a trend similar to that obtained for the consumption speed S$_{cF}$. However, the values of the stretched flame speed are generally slightly lower when S$_{c{\dot{Q}}}$ is used.

The evolution of S$_c$/$S_L^0$ in the aCFPF can be split into two main regions, referred to as the low-Ka region and the high-Ka region in the following. In the low-Ka region, the consumption speed varies differently with respect to strain depending on the fuel-air mixtures. It decreases for most of the mixtures with equivalence ratio of 0.9 but increases for mixtures with lower equivalence ratio. On the contrary, in the high-Ka region the consumption speed monotonically decreases as the strain rate increases until the flame reaches a condition of weak combustion (quasi-extinction). However, the flame never extinguishes completely as the heat coming from the product stream sustains some reactions even at the highest strain rates \cite{Dixon91}. It is seen in Fig.~\ref{plot_SRvSc_Ad}(a) that this quasi-extinction condition leads to a value of consumption speed closer to zero when the equivalence ratio decreases due to the lower temperature $T_{ad}$ of the product stream. 

An alternative version of an adiabatic counterflow premixed flame is obtained when the stagnation point flow is produced by two opposing streams of reactants, commonly known as twin flame (twinCFPF). This is equivalent to imposing a symmetry boundary condition in the stagnation point in Fig.~\ref{fig_CFPF_scheme}(a). Figure~\ref{plot_SRvSc_Ad}(b) presents the evolution of the consumption speed in the twinCFPF as a function of the strain rate. In the case of the twinCFPF, there is no high-Ka region because the flame experiences an abrupt extinction. This extinction is produced by the reduction in the residence time while the flame is pushed against the stagnation point \cite{Law89}. The maximum strain rate in the twinCFPF configuration is known as the extinction strain rate of the mixture. This extinction strain rate is also plotted in Fig.~\ref{plot_SRvSc_Ad}(a) for comparison. When the flame is pushed against the stagnation point in the aCFPF configuration, the reaction zone extends through the stagnation point to the product side. Then the reactions continue, eventually reaching the quasi-extinction condition.  
\color{black}


Figure~\ref{plot_SRvSc_Ad} highlights the influence of hydrogen addition and equivalence ratio on the response of a premixed CH$_4$-H$_2$-air laminar flame to positive strain. As reported in previous studies \cite{JacksSai03,Chen09a,SankaIm06,SpethMarzo05}, adding hydrogen to a lean methane flame not only increases the value of $S_L^0$, but also makes the flame respond positively to positive stretch, increasing the flame speed when the strain rate increases in the low-Ka region. This trend is easily seen for $\phi$=0.7 in Fig.~\ref{plot_SRvSc_Ad}(a), where the consumption speed remains nearly constant in the low-Ka region when the fuel is pure methane but increases for the CH$_4$-H$_2$ mixtures. The equivalence ratio significantly impacts the effect of H$_2$ on the response to strain. For the leaner condition, $\phi$=0.5 in Fig.~\ref{plot_SRvSc_Ad}, the consumption speed increases even for pure methane, and the effect of hydrogen addition is more pronounced. On the other hand, this effect is less pronounced when the mixture is closer to the stoichiometric condition.

\begin{figure}[h]
    \minipage{0.60\textwidth}
      \includegraphics[trim=130 290 140 300, clip, width=\linewidth]{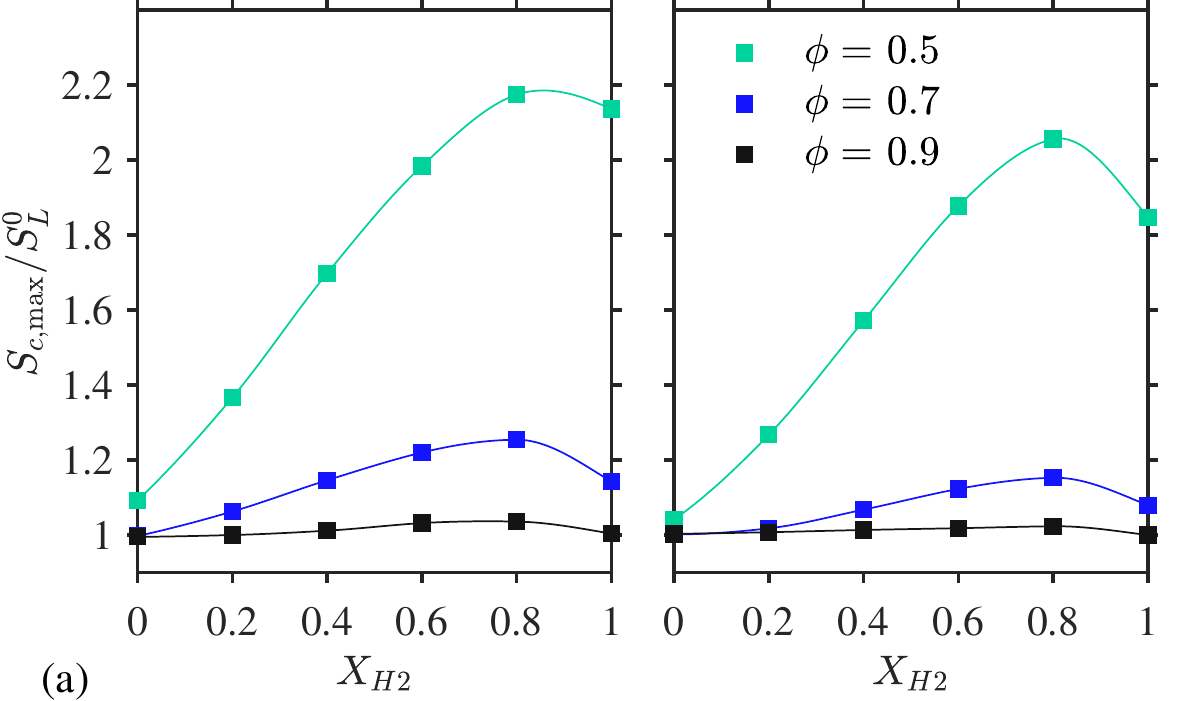}
    \endminipage
    \minipage{0.338\textwidth}
      \includegraphics[trim=203 290 215 300, clip, width=\linewidth]{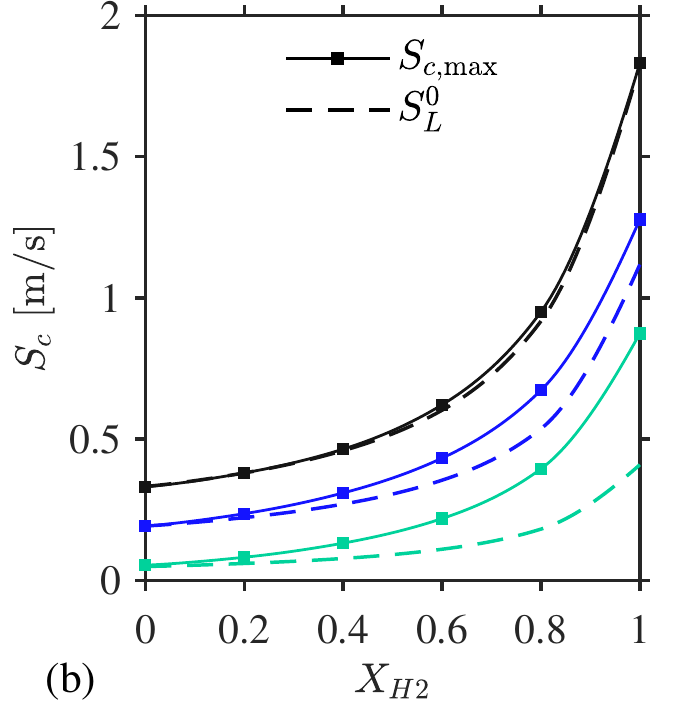}
    \endminipage
    \caption{(a) Normalized maximum stretched consumption speed, (left) based on fuel consumption S$_{cF}$, and (right) based on heat release rate S$_{c{\dot{Q}}}$. (b) Maximum stretched consumption speed S$_{cF}$ compared with unstreched laminar flame speed $S_L^0$.}
    \label{fig_ScMax} 
\end{figure}

Figure~\ref{fig_ScMax}(a) shows the evolution of maximum consumption speed in the adiabatic aCFPF as a function of the hydrogen concentration for various equivalence ratios using both definitions of the consumption speed. S$_{c,\mathrm{max}}/S_L^0$ increases as the equivalence ratio decreases and as the hydrogen addition increases up to $X_{H_2}$=0.8. The values of S$_{c,\mathrm{max}}$ for the CH$_4$-H$_2$ mixtures obtained in the aCFPF are lower than the ones obtained in the twinCFPF (as seen in Fig.~\ref{plot_SRvSc_Ad}). The difference is because in the aCFPF configuration the enhanced diffusion of radicals out of the reaction zone toward the product side of the flame plays an essential role, while in the twinCFPF, the reduced residence time is the only mechanism producing a reduction in the flame speed for Le$<$1.


The impact of the different responses of the flame speed to hydrogen addition can be observed by plotting the maximum consumption speed together with the unstretched laminar flame speed as shown in Fig.~\ref{fig_ScMax}(b). These results illustrate how various CH$_4$-H$_2$-air mixtures with the same $S_L^0$ could have different flame speeds along a turbulent premixed flame due to stretch induced by strain rate.

\subsection{Thermo-diffusive properties of the laminar flame}

\begin{figure}[h]
    \minipage{0.32\textwidth}
      \includegraphics[trim=203 286 215 300, clip, width=\linewidth]{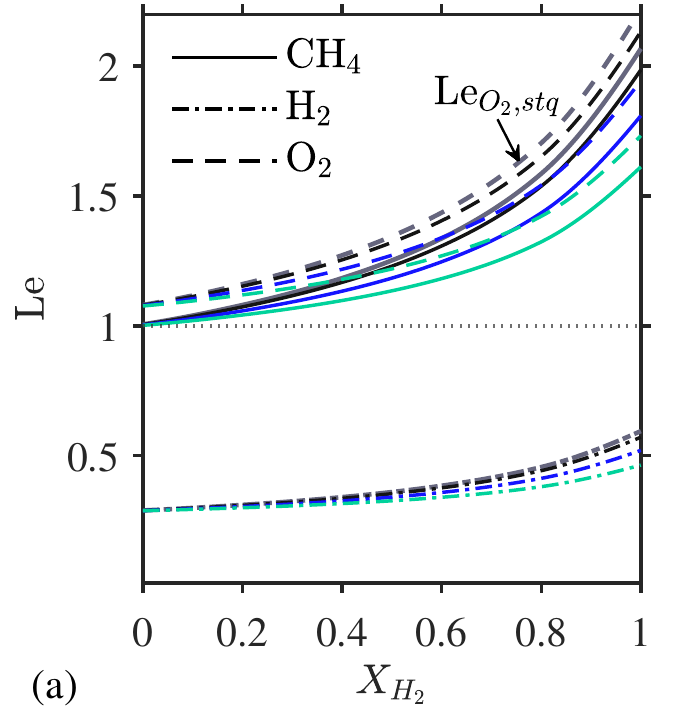}
    \endminipage
    \minipage{0.32\textwidth}
      \includegraphics[trim=203 286 215 300, clip, width=\linewidth]{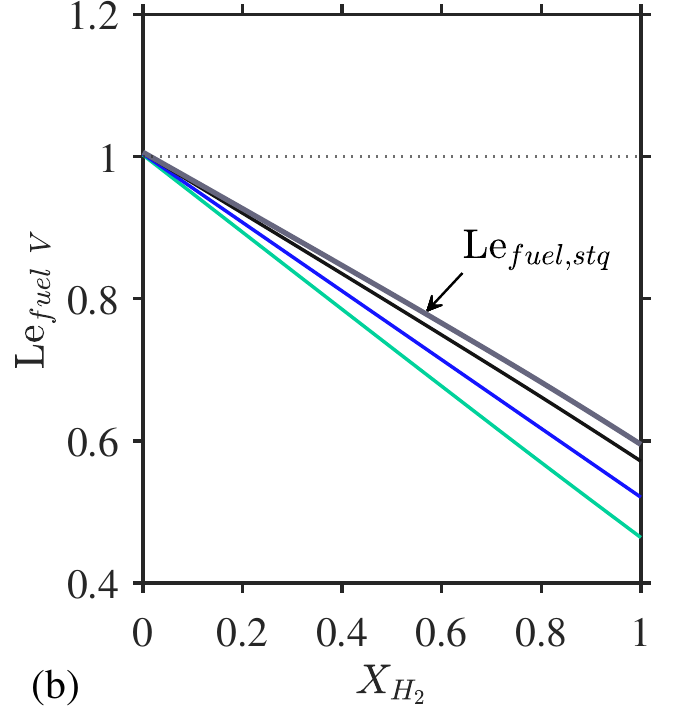}
    \endminipage
    \minipage{0.32\textwidth}%
      \includegraphics[trim=203 286 215 300, clip, width=\linewidth]{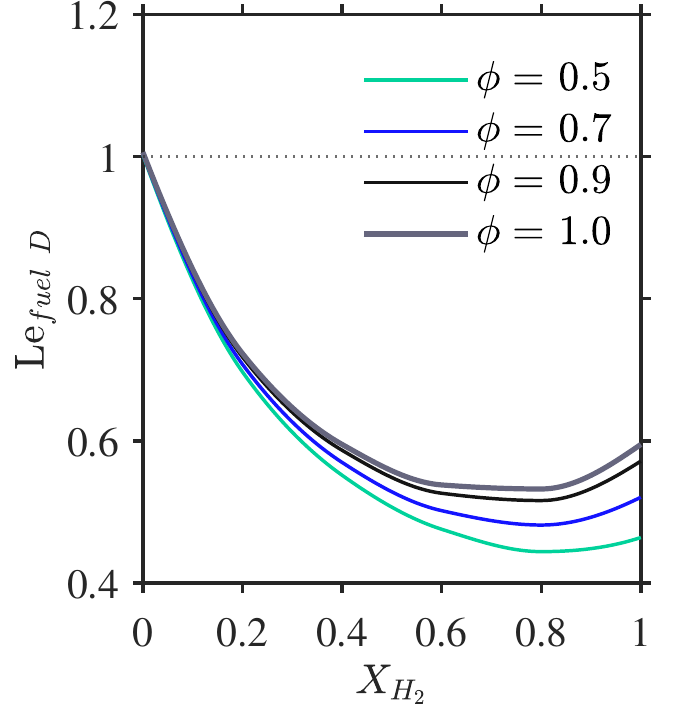}
    \endminipage
    \caption{(a) Lewis number of the reactants, (b) Lewis number of the fuel mixture CH$_4$-H$_2$, (left) based on the volume fraction and (right) based on the mass diffusion.}
    \label{fig_Le} 
\end{figure}

The change in the consumption speed with strain rate is a result of the variation of the convective-diffusive equilibrium in the flame front. Therefore, the definition of the consumption speed should agree with the thermal-diffusion properties of the fuel mixture. The thermal-diffusion of the CH$_4$-H$_2$-air mixtures are discussed in this section. First, Figure~\ref{fig_Le}(a) shows the Lewis number of each reactant species in the mixture for the various values of $X_{H_2}$ and $\phi$ values. The Lewis number for each species increases with $X_{H_2}$, due to the increase in the thermal diffusion of the mixture. The opposite happens with equivalence ratio. The thermal diffusion of the mixture, and therefore the Lewis number of each species, decreases when the mixture becomes leaner, especially for high hydrogen addition.

The Lewis number of the fuel mixture is presented in Fig.~\ref{fig_Le}(b) using the two definitions, Eqs~\eqref{Eq_LeV_fuel} and \eqref{Eq_LeD_fuel}. Both definitions give a $\mathrm{L e}_{fuel}$ varying from pure methane to pure hydrogen but with a different trend. The definition based on the mass diffusion varies more sharply for low hydrogen concentrations while varying non-monotonically when approaching pure hydrogen. In both cases $\mathrm{L e}_{fuel}$ is below unity for the CH$_4$-H$_2$ mixtures. For Le $<$ 1, positive strain rate increases the consumption speed because the enthalpy gain by reactant mass diffusion is greater than the heat loss by conduction from the reaction zone away from the reactants ahead \cite{Law89, LibbyLinn83}. The significant difference between the Lewis number of the hydrogen compared with the other reactants is due to the high mass diffusion of the former, which is around 3.5 times higher. A differential diffusion coefficient can characterize the difference in the mass diffusion of the various reactants as  $DD_{i-j} = \mathrm{L e}_i/\mathrm{Le}_j = D_{j-N_2}/D_{i-N_2}$. For positive strain and $DD_{fuel-O_2}<$1, the fuel diffuses toward the reaction zone more than the oxygen. This differential diffusion produces a richer mixture locally ahead of the reaction zone, which increases the flame speed for lean mixtures.

\begin{figure}[h]
    \minipage{0.32\textwidth}
      \includegraphics[trim=203 286 215 300, clip, width=\linewidth]{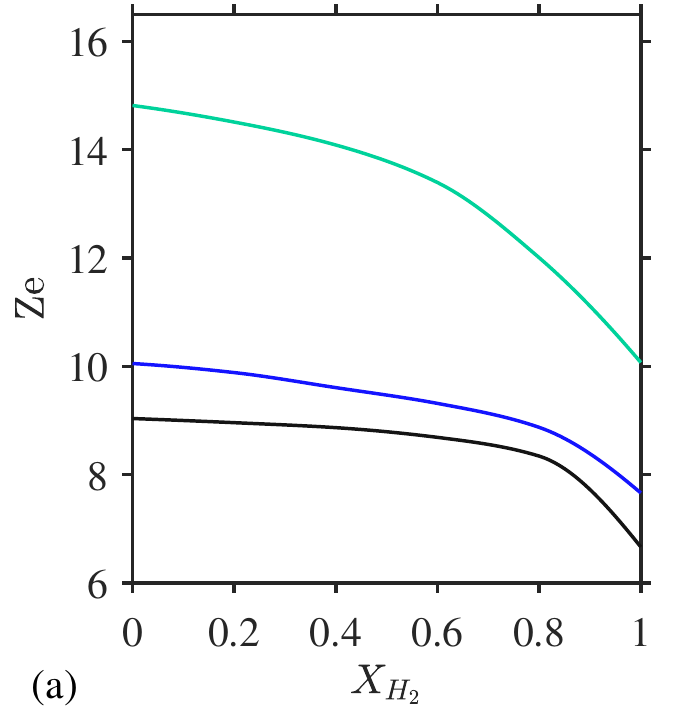}
    \endminipage
    \minipage{0.32\textwidth}
      \includegraphics[trim=203 286 215 300, clip, width=\linewidth]{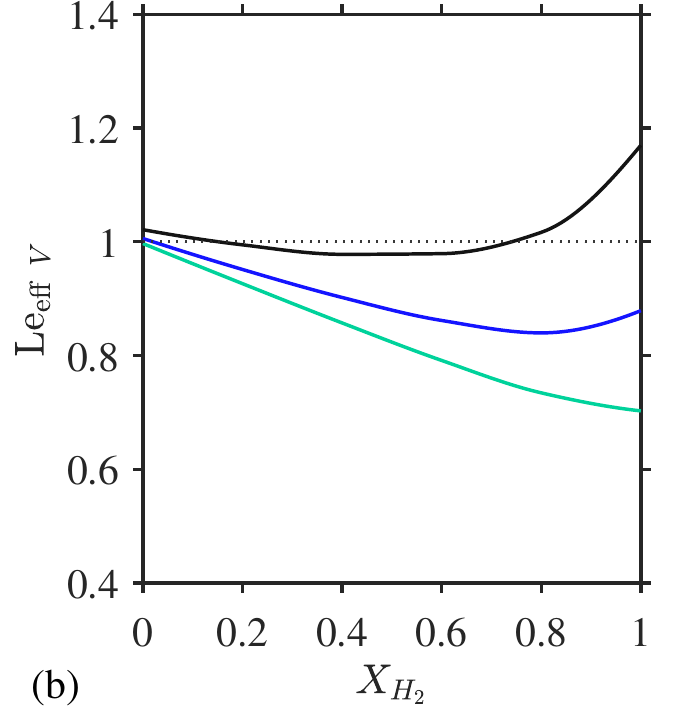}
    \endminipage
    \minipage{0.32\textwidth}%
      \includegraphics[trim=203 286 215 300, clip, width=\linewidth]{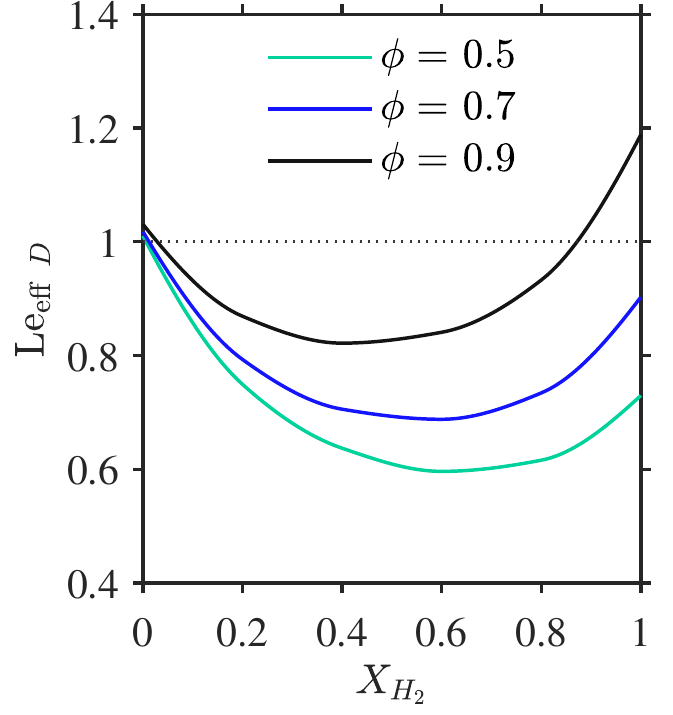}
    \endminipage
    \caption{(a) Zel’dovich number, and (b) effective Lewis number of the CH$_4$-H$_2$-air mixture, (left) based on the volume fraction and (right) based on the mass diffusion.}
    \label{fig_LeZe} 
\end{figure}

The effective Lewis number is used to consider both preferential diffusion effects, non-unity Lewis number and differential diffusion, in the fuel-air mixture. In Eq.~\eqref{Eq_Le_eff}, Le$_{\mathrm{eff}}$ is calculated using the Zel'dovich number, which characterizes the sensitivity of chemical reactions to the variation of the maximum flame temperature \cite{GuHaq00,BechtMatal01}. Figure~\ref{fig_LeZe}(a) presents the evolution of the Zel'dovich number of the fuel-air mixture as a function of $X_{H_2}$ for the various equivalence ratios. As observed in previous works \cite{SankaIm06,OkafoHayak14}, the Zel'dovich number decreases when hydrogen addition increases for all the equivalence ratios. This result is related to the higher reactivity of the hydrogen, which results in a mixture with lower overall activation energy. The equivalence ratio has the same effect on the Zel'dovich number. The latter increases when the mixture becomes leaner due to the increase of activation energy and the decrease of adiabatic flame temperature.

Figure~\ref{fig_LeZe}(b) shows the effective Lewis number as a function of $X_{H_2}$ and equivalence ratio based on the two definitions of $\mathrm{L e}_{fuel}$. As expected, the impact of equivalence ratio is less significant for low-hydrogen addition as methane and oxygen have similar Le (or similar mass diffusion coefficient). According to the definition of the effective Lewis number, the reactant species in deficit has a predominant effect on Le$_{\mathrm{eff}}$. This predominance grows when the species in deficit reduces with respect to the species in excess, which means that Le$_{\mathrm{eff}}$ gets closer to the Lewis number of the fuel as the mixture becomes leaner. The most significant difference occurs for the pure H$_2$-air mixture. Figure~\ref{fig_LeZe}(b) shows that for $\phi$ = 0.9, the effective Lewis number is above 1.0 for some CH$_4$-H$_2$-air mixtures. This observation is consistent with the results presented in Fig.~\ref{plot_SRvSc_Ad}(a), where, for similar values of $X_{H_2}$, the consumption speed decreases when the strain rate increases in the low-Ka region, as expected for a mixture with Le$>$1.  

\begin{figure}[h]
    \centering\minipage{0.40\textwidth}
      \includegraphics[trim=190 275 200 290, clip, width=\linewidth]{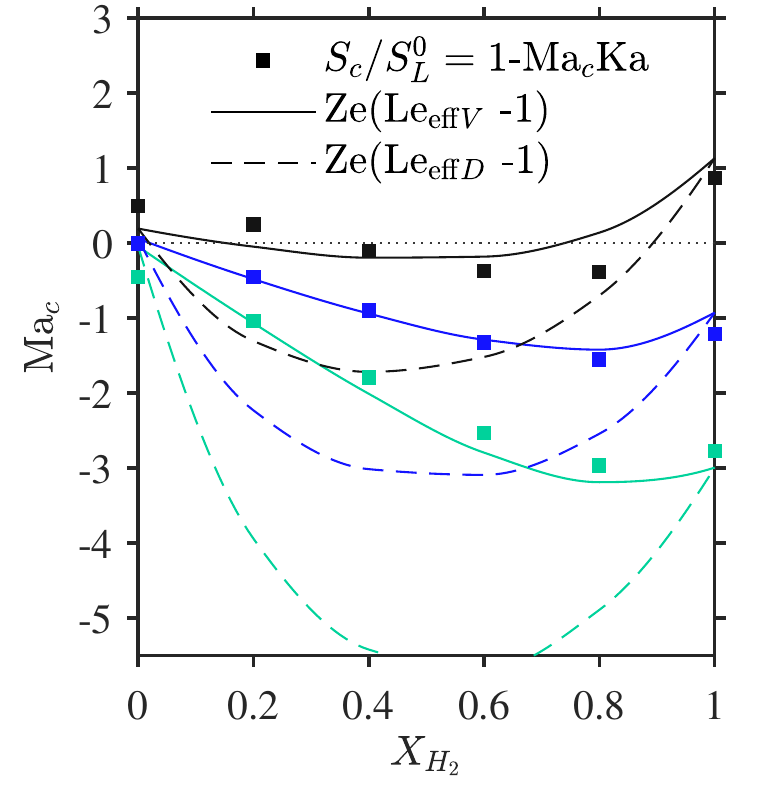}
    \endminipage
    \centering\minipage{0.40\textwidth}
      \includegraphics[trim=190 275 200 290, clip, width=\linewidth]{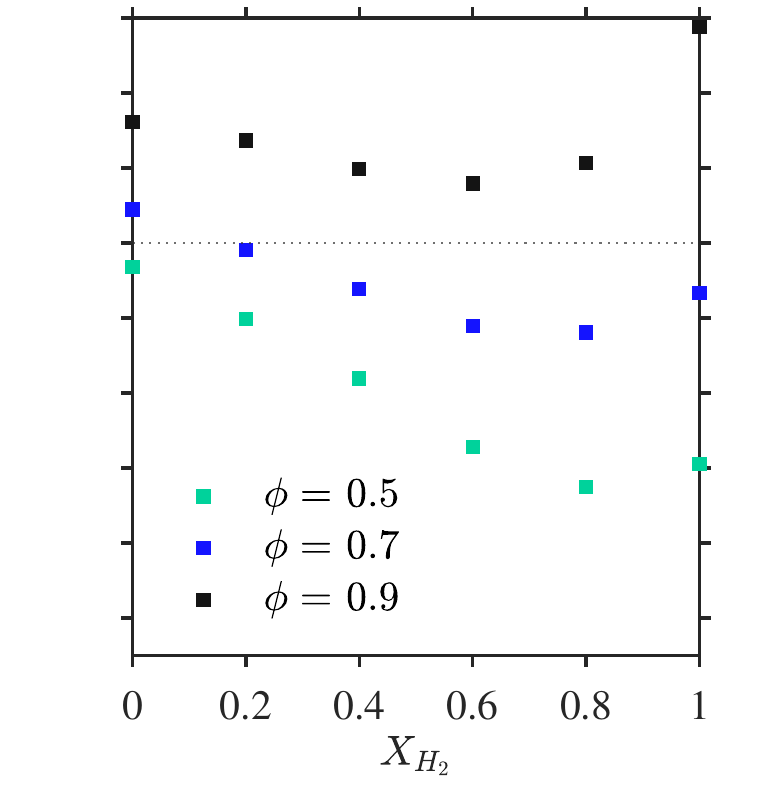}
    \endminipage
    \caption{Markstein number for the CH$_4$-H$_2$-air mixtures, (left) based on fuel consumption S$_{cF}$, and (right) based on heat release rate S$_{c{\dot{Q}}}$. The parameter $\mathrm{Ze}(\mathrm{Le}_{\mathrm{eff}}-1)$ is included to compare the definitions of $\mathrm{L e}_{fuel}$.}
    \label{fig_Ma_c} 
\end{figure}

For low  Karlovitz number, the relationship between the flame speed and the strain rate is linear and can be expressed as S$_c$/S$_L^0$=1-Ma$_c$Ka, where Ma$_c$ is the Markstein number based on the consumption speed \cite{BechtMatal01,DavisQuina02}. This number is a property of the fuel-air mixture and quantitatively describes the sensitivity of the flame speed to strain rate. Figure~\ref{fig_Ma_c} represents the evolution of Ma$_c$ with $X_{H_2}$ for both definitions of the consumption speed. The values of Ma$_c$ are determined by linear regression from the S$_c$/S$_L^0$ profiles presented in Fig.~\ref{plot_SRvSc_Ad}(a) for low strain rate range. 


The Markstein number varies non-monotonically with hydrogen addition, as previously observed experimentally in outwardly propagating spherical flames \cite{Chen09a, OkafoHayak14}. Adding hydrogen to a methane-air flame decreases Ma$_c$, but the trend is reversed for $X_{H_2}$ around 0.8, which means that adding methane to a hydrogen-air flame makes the diffusively unstable flame even more unstable, especially when the mixture becomes richer. The values of Ma$_c$ obtained with the definition of the consumption speed based on the heat release differ quantitatively, but not qualitatively, from those obtained with the definition based on fuel consumption. It seems that the effect of the equivalence ratio is more pronounced for Markstein numbers based on S$_{c{\dot{Q}}}$. For $\phi$ = 0.9, values of Ma$_c$ are always positive even for mixtures with effective Lewis number below one. This result highlights the influence of the consumption speed definition used to characterize the flame response to stretch and suggests that the definition based on the fuel consumption could better describe the speed of the CH$_4$-H$_2$ stretched flamelets.

It has been argued that $\mathrm{Ze}(\mathrm{Le}_{\mathrm{\mathrm{eff}}}-1)$ is a relevant parameter to describe the relationship between the flame response to strain rate and the thermo-diffusive properties of the mixture \cite{BechtMatal01, OkafoHayak14}. This parameter is computed using the two definitions of $\mathrm{L e}_{fuel}$ and the results are plotted Fig.~\ref{fig_Ma_c}. A remarkable agreement with the Ma$_c$ is obtained when the Lewis number of the fuel mixture is defined based on the volume fraction. In contrast, the definition based on the mass diffusion over-predicts the decrease in Markstein number of the CH$_4$-H$_2$-air mixtures, in disagreement with results from the aCFPF simulations.

\subsection{Consumption speed for strained laminar flame with heat loss}

\begin{figure}[h!]
    \centering\includegraphics[trim=0 185 0 190, clip, width=1\linewidth]{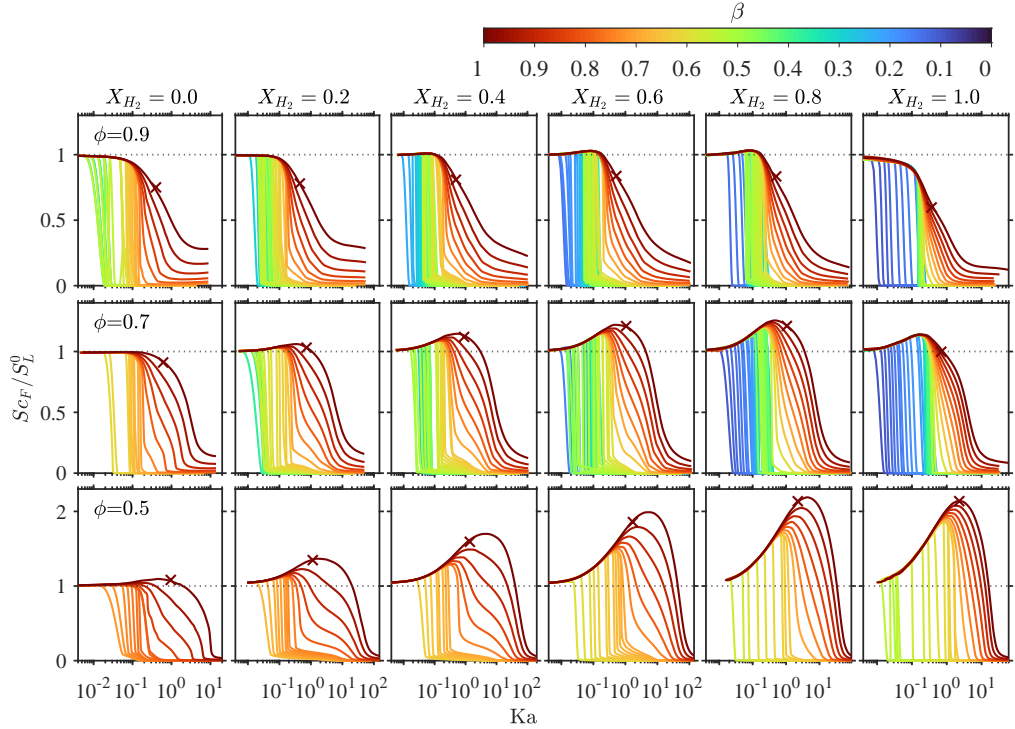}
    \caption{Response of consumption speed based on the fuel consumption, S$_{c_F}$, to the combined effect of strain and heat loss. The extinction strain rate from the twinCFPF is shown by the marker ($\times$) on top of the curve $\beta$ = 1.}
    \label{fig_SC_F-Ka} 
\end{figure}

Under non-adiabatic conditions, the flame experiences heat loss, which slows down the chemical reactions. This reduces the flame speed and can eventually cause extinction. The response of a premixed laminar flame to stretch is altered by heat loss, therefore, the propagation characteristics of the flame depend on its response to combined stretch and head loss. This section presents the effect of hydrogen addition on the response of the premixed laminar flame to combined positive strain rate and heat loss in the aCFPF configuration. Figure~\ref{fig_SC_F-Ka} presents the normalized consumption speed S$_c$/$S_L^0$ for various CH$_4$-H$_2$-air mixtures as a function of the Karlovitz number for values of $\beta$ ranging from 1 to 0. The response of the laminar flame to combined strain-heat loss behaves in two main ways depending on the level of heat loss, as reported in previous works for methane and propane laminar flames \cite{Law89,Dixon91,TangRaman21}. For low heat loss level ($\beta \simeq 1$), the consumption speed varies continuously with strain rate. For high heat loss level, however, the consumption speed sharply drops to reach a zero-value after the strain rate has exceeded a critical value that depends on the heat loss condition, but is always lower than the extinction strain rate obtained in the twinCFPF configuration, indicated by the marker $\times$ on top of the curve for $\beta$ = 1. The latter is perceived as an extinction event due to combined strain and heat loss, different from the quasi-extinction condition present for adiabatic or low heat loss levels. 

For low heat loss and strain rate  levels, the consumption speed varies independently of heat loss, following the same curve $\beta = 1$. This happens because, the flame is located far from the stagnation point; thus, the reaction zone is insulated by a diffusive-convective zone before the stagnation point. When the reaction zone approaches the stagnation point due to the increase of the strain rate, the heat loss adds to the negative effects of strain rate on the laminar flame, and it produces a faster decrease in the consumption speed. This combined effect of strain and heat loss results in a lower maximum stretched consumption speed for the flame under non-adiabatic conditions, especially for the case with the lowest equivalence ratio. The addition of hydrogen attenuates the impact of heat loss on the consumption speed response to strain rate, which can be seen in Fig.~\ref{fig_SC_F-Ka} as closer profiles of S$_c$/$S_L^0$ for a wider range of $\beta$.

The extinction of the flame due to combined strain and heat loss occurs for all the mixtures when the heat loss is high enough. The value of Karlovitz number at which extinction occurs varies highly non-monotonically with heat loss coefficient $\beta$. For 0.3$<\beta<$0.6, the extinction strain rate alternatively increases and decreases when the heat loss increases. Overall, however, the extinction tends to occur at lower strain rates when the heat loss increases. The trend continues until a condition of maximum heat loss is reached for which the flame is extinguished even at the minimum strain rate. The values of $\beta$ in Fig.~\ref{fig_SC_F-Ka} are varied by 0.05 initially and then by 0.01 to capture the behavior close to the extinction. The addition of hydrogen increases the resistance of the flame to combined strain and heat loss. The same happens with the equivalence ratio in agreement with its higher sensitivity to flame temperature variations, as seen in Fig.~\ref{fig_LeZe}(a). 

\begin{figure}[h!]
    \centering\includegraphics[trim=145 200 148 170, clip, width=0.5\linewidth]{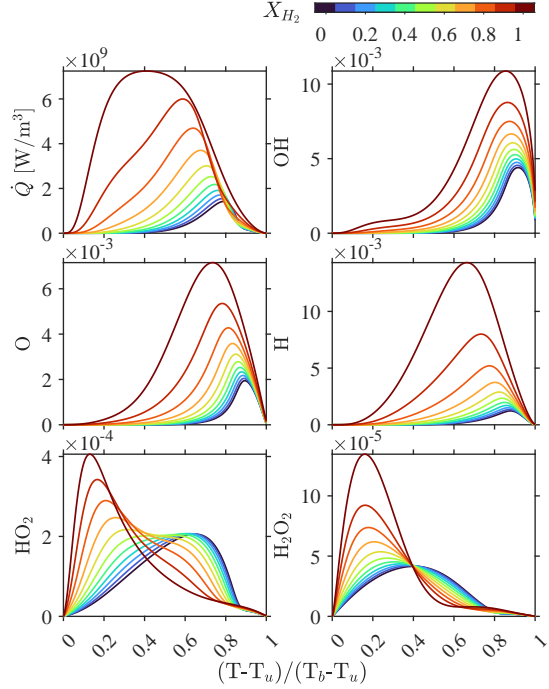}
    \caption{Effect of hydrogen addition on the internal structure of the laminar flame, shown by the profiles of the heat release rate and mole fraction of OH, O, H, HO$_2$, and H$_2$O$_2$ across the unstretched laminar flame thickness for $\phi=$0.7.}
    \label{fig_HRnRad_XH2} 
\end{figure}


Additionally to the decrease of the Zel’dovich number, the effect of hydrogen on the flame extinction by heat loss is related to the early oxidation of hydrogen. This increases the heat and radical production rates at low temperatures and allows methane oxidation under conditions where pure methane combustion might not be possible. To show this, Fig.~\ref{fig_HRnRad_XH2} presents the profiles of heat release and mole fraction of important radicals across the unstretched laminar flame with various hydrogen additions. The position through the flame front is indicated with the reduced temperature, which is of 0 and 1 at the unburnt and burnt sides of the flame, respectively. The reactions start at a lower temperature as the addition of hydrogen increases. This is illustrated by the location of the peak mole fraction of HO$_2$ and H$_2$O$_2$, which are two of the most important radicals controlling ignition and combustion at low temperature \cite{Westb00}.

\begin{figure}[h!]
    \centering\includegraphics[trim=0 185 0 190, clip, width=1\linewidth]{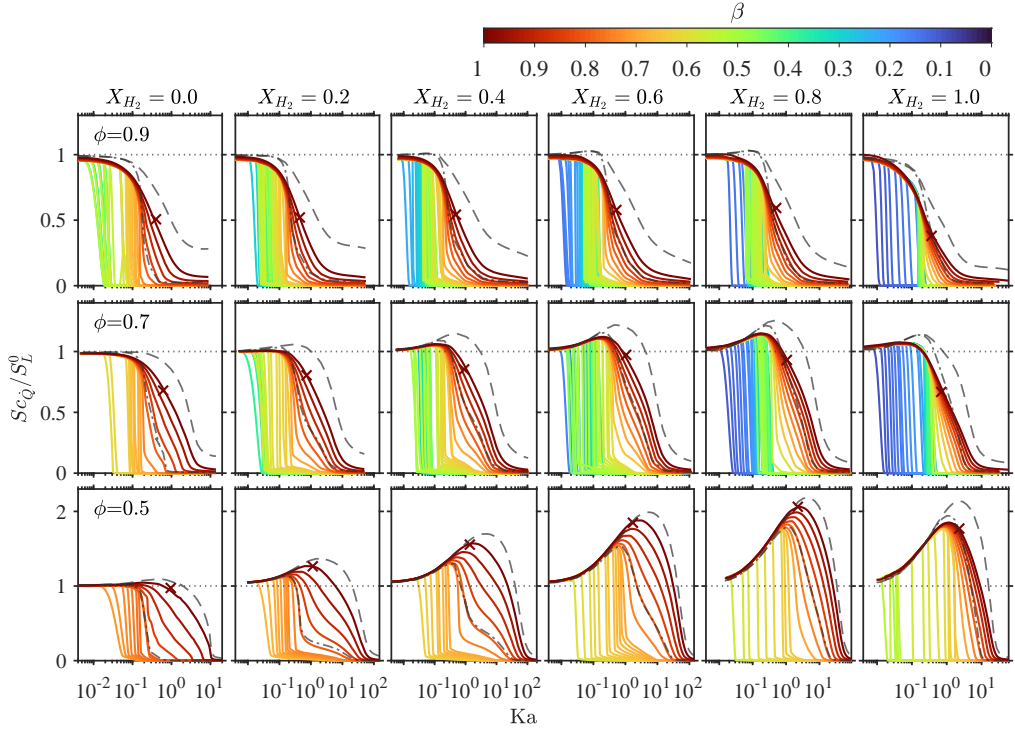}
    \caption{Response of consumption speed based on the heat release, S$_{c{\dot{Q}}}$, to the combined effect of strain and heat loss. The consumption speed based on the fuel consumption, S$_{cF}$, is indicated by dashed line for $\beta$ = 1.0 and a dash-dotted line for $\beta$ = 0.8. The extinction strain rate from the twinCFPF is shown by the marker ($\times$) on top of the curve $\beta$ = 1.}
    \label{fig_SC_HR-Ka} 
\end{figure}

The combined effect of strain rate and heat loss on the consumption speed based on the heat release rate is shown in Fig~\ref{fig_SC_HR-Ka}. Overall, the same qualitative trend is observed in the response to combined strain-heat loss. Quantitatively, the values of the stretched flame speed are lower compared with those obtained from the definition based on the fuel consumption rate. This difference is higher at the high-Ka region for the adiabatic condition, but diminishes as the heat loss level increases. This is shown in Fig~\ref{fig_SC_HR-Ka} where the values of S$_{cF}$ are plotted for $\beta$ = 1.0 and 0.8.

The discrepancy between the two definitions of the consumption speed can be explained by comparing the main reactions involved in the processes of fuel consumption and heat release. For the lean CH$_4$-H$_2$-air mixtures, only five reactions represent more than $99\%$ of the fuel consumption process. These reactions are: CH$_4$ + OH $\Leftrightarrow$ CH$_3$ + H$_2$O, CH$_4$ + H $\Leftrightarrow$ CH$_3$ + H$_2$, and CH$_4$ + O $\Leftrightarrow$ CH$_3$ + OH for methane; and H$_2$ + OH $\Leftrightarrow$ H + H$_2$O, and H$_2$ + O $\Leftrightarrow$ H + OH for hydrogen. They occur in the inner layer of the reaction zone and are directly enhanced by the hydrogen diffusion from the reactants. Therefore, the effect of strain rate results from the balance between the diffusion of hydrogen into the reaction zone and the diffusion of radicals out of the reaction zone. On the other hand, heat release is a multi-step process where many reactions are involved, and the contribution of each reaction may change with the reactant mixture, the strain rate and the heat loss. Around 30$\%$ of the heat release, e.g., is produced by a combination of the reactions: CH$_3$ + O $\Leftrightarrow$ CH$_2$O + H, H + H$_2$O + O$_2$ $\Leftrightarrow$ H$_2$O + HO$_2$, CO + OH $\Leftrightarrow$ CO$_2$ + H, and H$_2$ + OH $\Leftrightarrow$ H + H$_2$O. Most of these reactions are indirectly enhanced by hydrogen diffusion but are highly affected by the diffusion of radicals out of the reaction zone. Consequently, the negative effect of the strain rate on the consumption speed is more significant when the definition based on the heat release is used. When the flame is stretched and close to the stagnation point (high-Ka region), the discrepancy between the two definitions of the consumption speed reduces with increasing $\beta$, because heat loss becomes the predominant negative effect on the consumption speed.

\begin{figure}[h]
    \centering\includegraphics[trim=35 280 40 270, clip, width=1\linewidth]{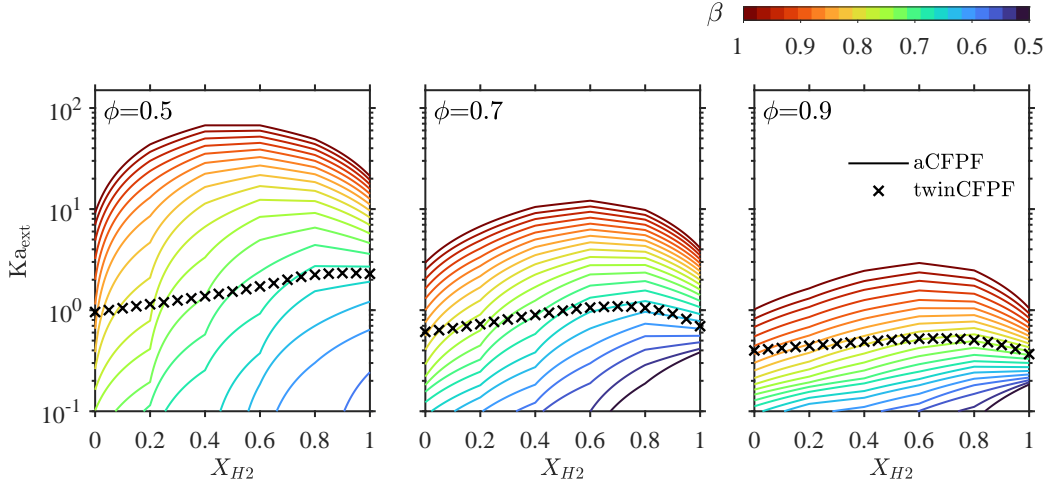}
    \caption{ Extinction Karlovitz number from asymmetric counterflow premixed flame with heat loss (Ka$_{\mathrm{ext-20\%\dot{Q}}}$). The extinction strain rate from the twinCFPF is shown by the marker ($\times$).}
    \label{fig_Ka_ext} 
\end{figure}

There is no general definition of the extinction strain rate in the aCFPF configuration due to the different extinction behaviors depending on the heat loss level. As mentioned above, a quasi-extinction condition is reached at high Ka for adiabatic or low heat loss conditions. To visualize the effect of hydrogen on this extinction behavior an extinction Ka is calculated based on the strain rate when the heat release-based consumption speed reaches 20$\%$ of the unstretched value. Figure~\ref{fig_Ka_ext} shows this extinction Karlovitz (Ka$_{\mathrm{ext}}$) for various CH$_4$-H$_2$-air mixtures as a function of the heat loss coefficient $\beta$ ranging from 1 to 0.5. The extinction strain rate computed in the twinCFPF is also shown for comparison. The latter is still defined as the maximum strain rate supported by the flame in this configuration. The extinction strain rate from the non-adiabatic aCFPF is higher than the one from the twinCFPF until the heat loss is high enough. The Ka$_{\mathrm{ext}}$ varies non-monotonically with hydrogen addition, especially for values of $\beta$ close to the adiabatic condition. The maximum Ka$_{\mathrm{ext}}$ occurs for mixtures with $X_{H2}$ between 0.4 and 0.8. When the heat loss increases ($\beta$ decreases), Ka$_{\mathrm{ext}}$ decreases more steeply with less hydrogen addition. Finally, the Ka$_{\mathrm{ext}}$ increases when the mixture becomes leaner due to the higher $\delta_L^0/S_L^0$ ratio (chemical time) used in the dimensionless strain rate, but also due to the increase in the preferential diffusion, which produces a stronger positive response to strain rate.

\section{Summary and conclusions}

A premixed counterflow flame in the asymmetric (aCFPF) configuration is used to study the effect of positive strain and heat loss on the consumption speed of lean methane-air laminar flames with hydrogen addition. A detailed chemical kinetic mechanism and full multi-component diffusion are used to describe the combustion process of the various mixtures, which range from pure methane to pure hydrogen, and three equivalence ratios to cover a wide range in the lean region. 
It is observed that the addition of hydrogen has a significant and non-monotonic effect on the response of the consumption speed to positive strain rate. This effect highly depends on the equivalence ratio of the fuel-air mixture and becomes more pronounced while the mixture gets leaner. The tolerance to both strain rate and heat loss is increased by hydrogen addition, which increases the maximum heat loss endured by the laminar flame before extinction due to the early oxidation of the hydrogen at lower temperatures. 

The maximum stretched consumption speed is reported for the lean CH$_4$-H$_2$-air mixtures. It can be easily seen from these results that mixtures with more H$_2$ addition but the same laminar flame speed can have higher turbulent flame speeds due to a higher maximum stretched consumption speed. This is relevant to modeling and understanding the propagation of turbulent flames with hydrogen. It is also shown that the values of the maximum stretched consumption speed obtained with the aCFPF configuration are lower than those obtained with the twinCFPF configuration due to a better representation of the boundary conditions prevailing in a realistic flame front, where the diffusion of radicals from the reaction zone to the burnt side of the flame plays an important role on the response to stretch.  

Two definitions of the consumption speed are compared, showing that different results are obtained. The definition based on the heat release leads to lower values of the flame speed under positive strain rate and heat loss than the definition based on the fuel consumption. This behavior results from a more significant impact of strain rate and heat loss on the multiple reaction steps involved in the heat release process. Therefore, the two different expressions of the consumption speed are expected to yield the same value in the context of one-step chemistry and begin to diverge while the heat release process is distributed over more reaction steps in a detailed chemistry.

The response of the laminar flame to strain rate and heat loss is related to the effect of hydrogen on the thermo-diffusive properties of the fuel-air mixture, mainly described by the Zel’dovich number and the effective Lewis number. The definition based on the fuel consumption is found to agree with the thermo-diffusive properties of the mixture for the range of equivalence ratio and hydrogen addition evaluated. On the other hand, the definition based on the heat release exhibits a trend opposite to that of the thermo-diffusive properties for mixtures close to the stoichiometric equivalence ratio. This result suggests that the former definition could better describe the speed of the CH$_4$-H$_2$ premixed flamelets exposed to turbulence-induced stretched and heat loss. Comparing both definitions in the context of turbulent flame would confirm this conclusion. 

Two commonly used approaches to compute the Lewis number of the fuel mixture are also compared. The results obtained with the definition based on the volume fraction are in better agreement with the simulations of the aCFPF than those obtained with the definition based on the mass diffusion.

The results of the present work can be used to understand and explain the effect of hydrogen addition on the shape and dynamics of a flame in real combustion systems exposed to high levels of aerodynamic strain and heat loss, a scenario that is relevant for many applications. Additionally, the results presented are relevant in the context of turbulent combustion modeling of CH$_4$-H$_2$-air flames. On the one hand, kinematic models used to track the flame front, like the G-Equation, or to calculate the flame propagation, like the flame-surface-density models, rely on the laminar flame speed that is normally fixed constant along the flame front. In the cases of H$_2$ enriched flames, the effect of stretch must be considered. This can be included directly in the laminar flame speed or by an additional stretch model. On the other hand, these aCFPF simulations can be used as a reference to validate reduced chemical mechanisms or to build a tabulated chemistry for flamelet-based models where the effect of heat loss and strain rate are included as trajectories in the model table.


\section*{Acknowledgement(s)}

\noindent
\begin{minipage}{0.25\textwidth}
\includegraphics[trim=5.5cm 10cm 4cm 11cm, clip,width=\linewidth]{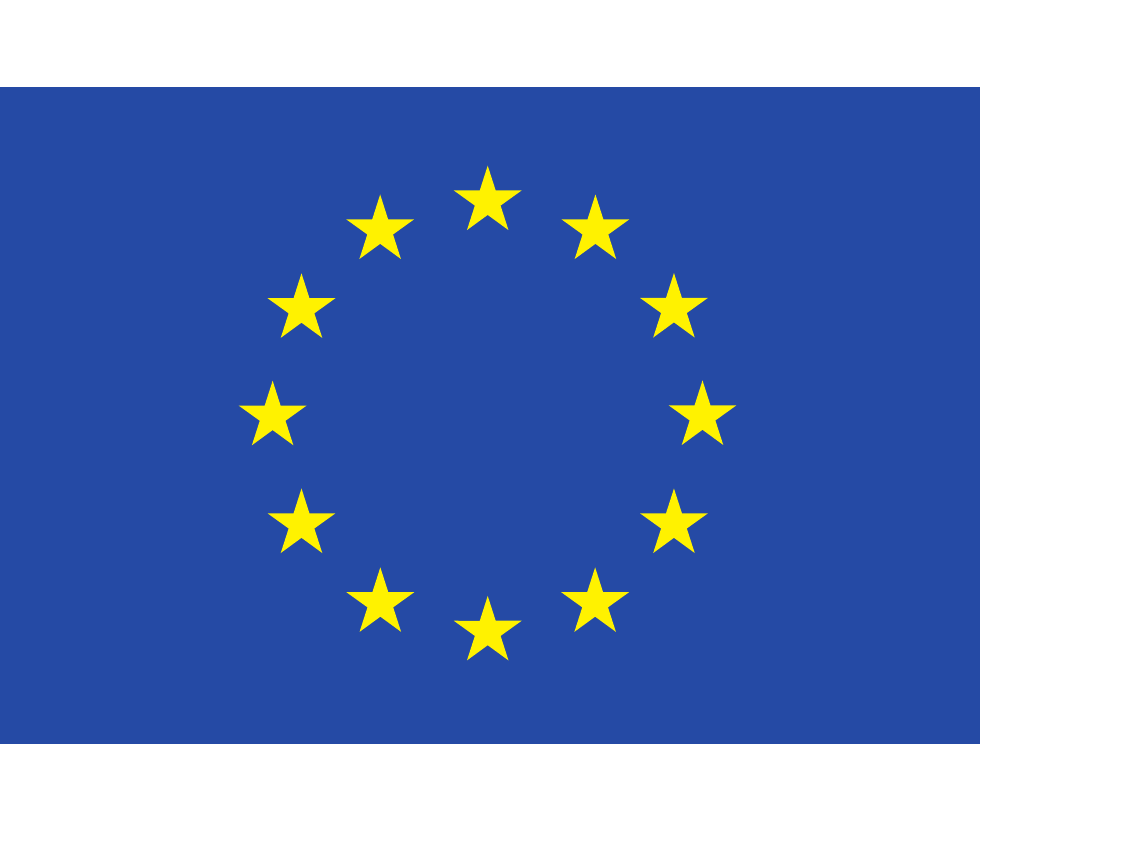}
\end{minipage}
\begin{minipage}{0.75\textwidth}
This work is part of the Marie Skłodowska-Curie Innovative Training Network Pollution Know-How and Abatement (POLKA). We gratefully acknowledge the financial support from the European Union’s Horizon 2020 research and innovation programme under the Marie Skłodowska-Curie grant agreement No. 813367.
\end{minipage}

\section*{Author Declarations}
The authors report there are no competing interests to declare.

\bibliographystyle{tfq}
\bibliography{Biblio.bib}

\end{document}